\documentclass{Iberspeech2026}

\interspeechcameraready 

\title{Disentangling Speaker and Language Effects in Cross-Lingual \\Speaker Verification for Iberian Languages}

\author[affiliation={1,2,*}, orcid=0009-0005-1055-129X]{Pol}{Buitrago}
\author[affiliation={2}, orcid=0000-0002-1730-8154]{Javier}{Hernando}

\affiliation{Barcelona Supercomputing Center (BSC)}{Spain}
\affiliation{Universitat Politècnica de Catalunya (UPC)}{Spain}
\email{pol.buitrago@bsc.es}

\usepackage{comment}
\usepackage{soul}

\begin{document}

\maketitle

\begingroup
\renewcommand{\thefootnote}{\fnsymbol{footnote}}
\footnotetext[1]{Corresponding author} 
\endgroup

\begin{abstract}
Cross-lingual speaker verification (SV) systems typically exhibit performance degradation when enrollment and test utterances are spoken in different languages. However, standard evaluation protocols confound language mismatch with inter-speaker variability, as evaluation is generally performed with different speakers across languages.
    
In this work, we introduce a bilingual same-speaker evaluation set for five Iberian languages, enabling analysis of cross-lingual SV under constant speaker identity. We apply this setup to a HuBERT-based SV system previously shown to exhibit strong language dependence, and analyze results using the Cross-Lingual Transfer Matrix (CLTM) to study pairwise cross-lingual transfer.
    
Our results show that speaker-related variability accounts for part of the observed degradation, but language mismatch remains the main driver of cross-lingual performance loss. These findings provide a more precise characterization of language dependence in cross-lingual SV.
\end{abstract}

\section{Introduction}

Speaker Verification (SV) systems aim to determine whether two utterances originate from the same speaker by extracting embeddings that capture identity-specific traits. Because speaker identity is largely independent of linguistic content, SV is expected to rely primarily on extralinguistic cues and to generalize across languages \cite{Winters2008,menon2025laspalanguageagnosticspeaker,9522962}. However, multiple studies have shown that cross-lingual mismatch can lead to substantial performance degradation \cite{940862,li2017crosslingualspeakerverificationdeep,buitrago2026cltransfer_arxiv}, with systems trained on one language often performing significantly worse when evaluated on another \cite{9522962, Chen2020}. These results suggest that, while SV is commonly considered a paralinguistic task, the acoustic representations learned by current models are not fully language-agnostic. 


This effect has been observed across a wide range of architectures and settings \cite{reuter2025language,1327105,buitrago2026towards}, and can sometimes be mitigated through specific modeling choices \cite{1327105,7078603,7919004,Chen2020}. In our previous work, we analyzed multiple distinct setups and showed that the degree of cross-lingual degradation in SV depends strongly on the underlying architecture \cite{buitrago2026towards}, indicating that some systems learn more language-robust representations than others.

However, while cross-lingual degradation can be characterized and quantified across systems, it remains unclear to what extent it is attributable to language mismatch rather than other factors. In standard evaluations, cross-lingual performance is measured using different speakers across languages due to the difficulty of collecting large-scale corpora with the same individuals speaking multiple languages. Consequently, language effects are confounded with speaker variability in existing benchmarks, preventing clear attribution of the observed degradation to language differences alone \cite{9746210}.

Previous studies using bilingual speakers have shown that cross-lingual SV remains sensitive to language mismatch even under same-speaker evaluation conditions \cite{Winters2008,7078603,90cd53ff49bf4a3fac11180e625ca6cb}. This degradation has been attributed to domain shift induced by differences in spoken language, even when produced by the same speaker \cite{9522962}. These studies suggest that, while language mismatch persists under same-speaker evaluations, speaker variability introduces an additional source of variation on top of language mismatch in cross-lingual SV. However, these findings do not allow a clear assessment of the respective contributions of speaker-related and language-dependent factors.

To address this underexplored aspect, we focus on five Iberian languages, namely Spanish, Catalan, Galician, Basque, and Portuguese, as a case study. This set of languages provides an interesting testbed for disentangling speaker and language effects due to their shared historical roots, high lexical similarity, and varying degrees of phonetic overlap. We use this setting to analyze cross-lingual behavior in a HuBERT-based SV system previously shown to exhibit strong language dependence even for closely related languages \cite{buitrago2026cltransfer_arxiv}.

To study the confounding between speaker variability and language-specific effects in standard cross-lingual evaluation, we construct a set of bilingual speaker test conditions using the Common Voice corpus \cite{ardila2020commonvoicemassivelymultilingualspeech}. Specifically, we select speakers who have recorded utterances in Spanish and at least one additional Iberian language, and organize them into bilingual subsets for each language pair. We compare this bilingual same-speaker evaluation against the standard protocol based on different speakers per test set, enabling a direct assessment of speaker factors in cross-lingual performance.

To analyze cross-lingual transfer across all language pairs, we use the Cross-Lingual Transfer Matrix (CLTM) \cite{buitrago2026cltransfer_arxiv}, a performance-grounded framework for quantifying pairwise cross-lingual transfer. The CLTM measures how adding training data from a donor language during fine-tuning affects performance on a target language relative to an equivalent amount of target-language data, enabling a systematic and quantitative characterization of cross-lingual transfer effects.

The aim of this work is threefold: (i) to introduce a bilingual same-speaker test set for cross-lingual SV in Iberian languages; (ii) to apply the CLTM framework to analyze cross-lingual transfer under both standard and same-speaker evaluation settings; and (iii) to analyze the extent to which speaker variability affects cross-lingual SV degradation, while showing that language mismatch remains the dominant factor.

\section{Method}
\label{sec:method}


We analyze cross-lingual transfer in SV using our previously proposed Cross-Lingual Transfer Matrix (CLTM), a framework that quantifies how incorporating training data from a donor language affects target-language performance relative to an equivalent amount of target-language data. A brief summary of the formulation is provided below; full details can be found in \cite{buitrago2026cltransfer_arxiv}.

\subsection{Cross-Lingual Transfer Matrix (CLTM)}
\label{sec:cltm}

Let $D_i$ and $D_i'$ denote disjoint training subsets from target language $i$, and $D_j$ from donor language $j$, each containing $N$ samples. We denote by $\mathrm{Perf}_i(\cdot)$ the performance evaluated on language $i$. Performance gains from additional target (\textit{self-gain}) and donor data (\textit{cross-gain}) are defined as:\vspace{-0.05cm}
\begin{equation}
\begin{aligned}
\Delta_{i\leftarrow i} &= \mathrm{Perf}_i(D_i + D_i') - \mathrm{Perf}_i(D_i),\\
\Delta_{i\leftarrow j} &= \mathrm{Perf}_i(D_i + D_j) - \mathrm{Perf}_i(D_i).
\end{aligned}
\end{equation}

The CLTM is defined as the normalized ratio
\begin{equation}
\mathrm{CLTM}[i,j] = \frac{\Delta_{i\leftarrow j}}{\Delta_{i\leftarrow i}},
\end{equation}
enabling cross-lingual comparison of donor effects.\vspace{0.2cm}

Each entry reflects the impact of donor language $j$ on target language $i$ relative to an equivalent amount of target-language data. Negative values indicate negative transfer, values in $(0,1)$ indicate indicate positive but weaker transfer than target-language data, and values above $1$ indicate stronger transfer than the corresponding target-language contribution.

A fully language-agnostic task, where all languages contribute equally to the others, would yield $\mathrm{CLTM} = \mathbf{1}_{n\times n}$.\vspace{0.2cm}

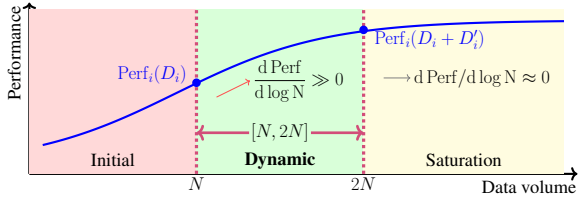
\begin{figure}[h!]
\centering
\resizebox{0.95\linewidth}{!}{%
\begin{tikzpicture}[x=2.2cm,y=0.9cm, every node/.style={font=\LARGE}, transform shape]

  \pgfmathsetmacro{\xmin}{0}
  \pgfmathsetmacro{\xmax}{8.2}
  \pgfmathsetmacro{\ymin}{0}
  \pgfmathsetmacro{\ymax}{6}

  \coordinate (Xstart) at (\xmin,\ymin);
  \coordinate (Xend)   at (\xmax,\ymin);
  \coordinate (XplotEnd) at ({\xmax-0.25},\ymin);
  \coordinate (Yend)   at (\xmin,\ymax+0.25);

  \draw[->, line width=2pt] (Xstart) -- (Xend);
  \draw[->, line width=2pt] (Xstart) -- (Yend);

  \path (Xstart) -- (XplotEnd) coordinate[midway] (XmidPlot);
  \node[anchor=north, font=\LARGE] at ($(XmidPlot)+(3.5,-0.12)$) {Data volume};

  \path (Xstart) -- (Yend) coordinate (Ymid);
  \node[left=14pt, rotate=90, font=\LARGE] at (Ymid) {Performance};

  \fill[red!15] (0,\ymin) rectangle (2.5,\ymax);        
  \fill[green!15] (2.5,\ymin) rectangle (5.0,\ymax);    
  \fill[yellow!15] (5.0,\ymin) rectangle (8.0,\ymax);   

  \draw[very thick, line width=2pt, domain=0.2:8.0, smooth, samples=200, blue]
      plot (\x, {(\ymax-0.7)/(1+exp(-0.9*(\x-2.2))) + 0.3});

  \draw[dashed, purple!70, line width=3pt] (2.5,\ymin) -- (2.5,\ymax);
  \draw[dashed, purple!70, line width=3pt] (5.0,\ymin) -- (5.0,\ymax);
  \draw[<->, line width=2.5pt, purple!70] (2.5,1.5) -- (5.0,1.5); 
  \node[fill=green!15, rounded corners=2pt, inner xsep=2pt, inner ysep=3pt] 
      at (3.75,1.5) {$[N,2N]$};

  \draw[->, thick, red!70] (2.8,2.8) -- (3.3,3.4)
      node[right, black, font=\LARGE] {$\dfrac{\mathrm{d\,Perf}}{\mathrm{d\log N}} \gg 0$};
  \draw[->, thick, gray] (5.3,3.5) -- (5.7,3.5)
      node[right, black, font=\LARGE] {$\mathrm{d\,Perf}/\mathrm{d\log N} \approx 0$};

  \fill[blue] (2.5, {(\ymax-0.7)/(1+exp(-0.95*(2.5-2.2))) + 0.3}) circle (4pt)
      node[left, xshift=-6pt, yshift=8pt, font=\LARGE] {$\mathrm{Perf}_i(D_i)$};
  \fill[blue] (5.0, {(\ymax-0.7)/(1+exp(-0.95*(5.0-2.2))) + 0.3}) circle (4pt)
      node[right, xshift=7pt, yshift=-9pt, font=\LARGE] {$\mathrm{Perf}_i(D_i + D_i')$};

  \node[anchor=center, font=\LARGE] at (1.25,0.5) {Initial};
  \node[anchor=center, font=\LARGE\bfseries] at (3.75,0.5) {Dynamic};
  \node[anchor=center, font=\LARGE] at (6.5,0.5) {Saturation};

  \node[below, font=\LARGE] at (2.5,0) {$N$};
  \node[below, font=\LARGE] at (5.0,0) {$2N$};

\end{tikzpicture}}
\vspace{-0.2cm}
\caption{Typical learning curve for a single language, showing the dynamic interval and derivative regimes.}
\label{fig:learning_curve_clean}
\end{figure}

The CLTM is computed in a regime where performance variations due to additional training data are reliably measurable. We identify the interval $[N,2N]$ empirically from learning curves, selecting the region where performance increases significantly before saturation (the ``Dynamic'' region in Figure~\ref{fig:learning_curve_clean}). This ensures that both $\mathrm{Perf}_i(D_i)$ and $\mathrm{Perf}_i(D_i + D_j)$ fall within a regime where differences between conditions can be meaningfully observed.\vspace{-0.15cm}

\vspace{-0.1cm}

To characterize cross-lingual transfer, we use several metrics derived from the CLTM formulation:

\begin{itemize}
    \item {Relative Frobenius Deviation ($\mathrm{RFD}_1$):} quantifies the deviation of the CLTM from the language-agnostic case ($\mathbf{1}_{n\times n}$). Lower values indicate more language-agnostic behavior.

    \item {Relative Asymmetry ($\mathrm{Asym}_{\mathrm{rel}}$):} measures directional differences between donor and target; zero indicates symmetry.

    \item {Average Row Cosine Similarity ($\overline{\cos}_{\mathrm{rows}}$):} assesses similarity between transfer profiles across target languages, with higher values indicating more consistent donor effects.

    \item {Additional statistics:} derived from off-diagonal entries, we also consider the proportion of positive transfer (prop$_+$).
\end{itemize}

\section{Experimental Setup}
\subsection{Data}
\label{sec:data}

We use the Mozilla Common Voice corpus 25.0 \cite{ardila2020commonvoicemassivelymultilingualspeech}, which provides broad multilingual coverage of Iberian languages under a uniform recording protocol, reducing extraneous variability. Its large speaker pool enables the presence of multilingual speakers, essential for our bilingual evaluation setup.

For training, we construct strictly balanced language subsets with equal numbers of samples and disjoint speaker identities per language. Each speaker contributes a fixed number of 50 utterances, ensuring that performance differences cannot be attributed to dataset imbalance or speaker frequency effects.

For the bilingual same-speaker evaluation, we select language pairs with sufficient overlap of multilingual speakers: Spanish–Catalan (300 speakers), Spanish–Basque (64), Spanish–Portuguese (40), and Spanish–Galician (21). These speakers are paired across languages to form controlled same-speaker cross-lingual trials. We also retain the original Common Voice test partitions for standard cross-lingual evaluation.

For each target language $i$, performance $\mathrm{Perf}_i(D_i)$ is computed using $N$ samples, and $\mathrm{Perf}_i(D_i + D_i')$ after augmenting with a disjoint set of $N$ samples. Cross-lingual performance $\mathrm{Perf}_i(D_i + D_j)$ is computed for all pairs $(i,j)$ with $i \neq j$.

\subsection{Dynamic Training Interval}
\label{sec:dat_interval}

\begin{figure}[h!]
\centering
\includegraphics[width=\linewidth]{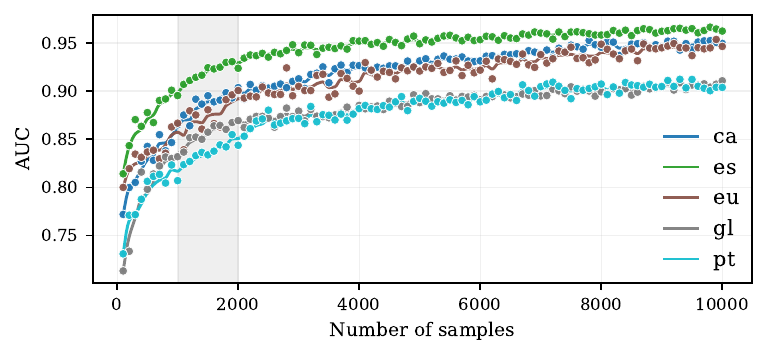}\vspace{-0.35cm}
\caption{Learning curves for the five target languages used to determine the dynamic training interval $[N,2N]$.}
\label{fig:learning_curve}
\end{figure}

The dynamic training interval is set to $[N,2N] = [1000,2000]$ samples per language, selected from preliminary learning curves to ensure positive self-gains ($\Delta_{i\leftarrow i} > 0$) across the five target languages (Fig.~\ref{fig:learning_curve}). This range avoids both undertraining and performance saturation, ensuring reliable CLTM estimates.

\vspace{-0.35cm}

\subsection{Model and Training}

\begin{figure*}[!b]
\centering\vspace{-0.3cm}
\includegraphics[width=0.9\textwidth]{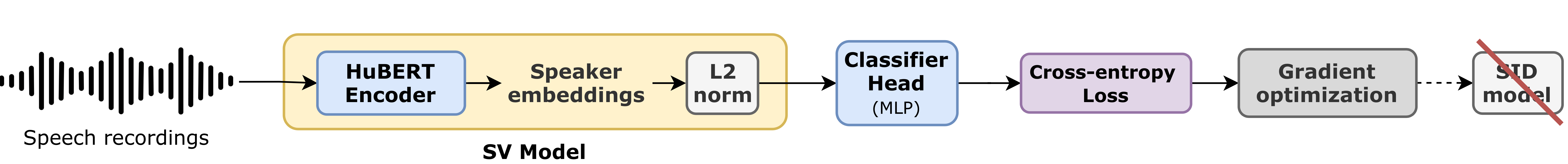}\vspace{-0.15cm}
\caption{Training pipeline of the HuBERT-based speaker verification system. The multilingual encoder is fine-tuned for speaker identification, after which the classification head is discarded and the encoder is used to extract speaker embeddings for verification.}
\label{fig:hubert}
\end{figure*}

We use the multilingual \texttt{mHuBERT-147}\footnote{Model checkpoint used in this work: \url{https://huggingface.co/utter-project/mHuBERT-147}} encoder \cite{boito2024mhubert147compactmultilingualhubert}, a HuBERT-based model pretrained on 147 languages, including all Iberian languages considered in this work. Following \cite{buitrago2026cltransfer_arxiv}, the encoder is fine-tuned for speaker identification by appending a randomly initialized linear classification head and optimizing a cross-entropy objective. After training, the classification head is discarded and the fine-tuned encoder is used as a speaker embedding extractor. The training procedure is illustrated in Fig.~\ref{fig:hubert}.

Audio is resampled to 16~kHz and processed using continuous encoder representations. All models are trained for a single epoch to minimize optimization-related confounds, using AdamW \cite{loshchilov2019decoupledweightdecayregularization} with a learning rate of $10^{-5}$, jointly updating all encoder parameters. To reduce the influence of data composition and initialization effects, each experiment is repeated across 20 independent data samplings and 5 random seeds, resulting in 100 runs per configuration. Reported results correspond to the average across all runs.

Embeddings are obtained by temporally pooling transformer hidden representations followed by L2 normalization. During evaluation, embeddings are computed independently for each utterance and compared using cosine similarity for SV. Performance is reported using the Area Under the Curve (AUC).


\vspace{-0.2cm}

\section{Results}
\label{sec:results}

We first compute the CLTM for all considered Iberian languages under the standard evaluation protocol, using it to characterize overall cross-lingual transfer patterns. We then compare these results against same-speaker evaluation conditions to assess the contribution of speaker variability to cross-lingual degradation.


CLTMs are visualized as heatmaps\footnote{The code required to reproduce model training and all analyses is available at \url{https://github.com/Pol-Buitrago/cltm-framework}.}, with dashed regions highlighting transfers within the Romance language group.

\subsection{Baseline Cross-Lingual Transfer Patterns}
\label{sec:baseline}

\begin{figure}[H]
    \centering\vspace{-0.4cm}
    \includegraphics[width=\columnwidth]{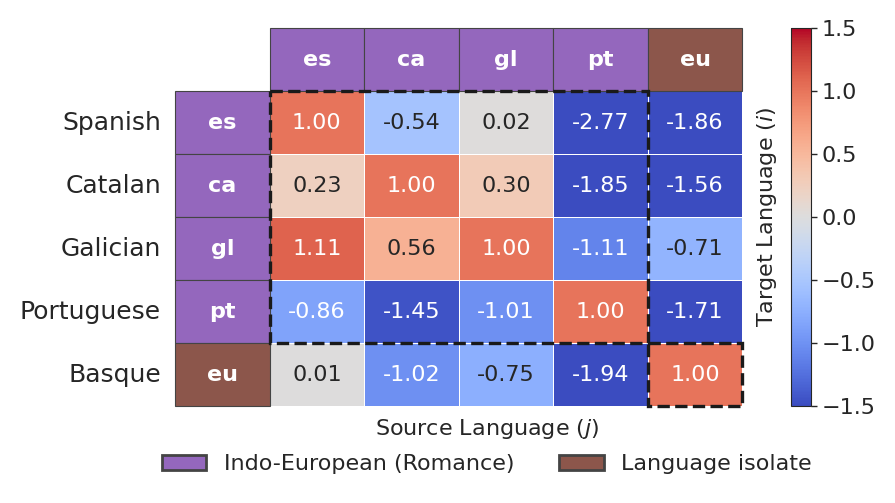}\vspace{-0.5cm}
    \caption{CLTM obtained under the standard evaluation.} 
    \label{fig:cltm_full}\vspace{-0.35cm}
\end{figure}

Fig.~\ref{fig:cltm_full} shows the 5$\times$5 CLTM for the five Iberian languages, obtained under the standard evaluation protocol. The matrix departs markedly from the language-agnostic ideal, with structured and highly asymmetric transfer patterns. 
Spanish, Catalan, and Galician form the main positive cluster, while Basque and Portuguese mostly yield negative transfer, indicating strong language dependence in the learned representations.

This is reflected in the aggregate diagnostics in Table~\ref{tab:baseline_cltm_diagnostics}. Compared to the full 44-language CLTM \cite{buitrago2026cltransfer_arxiv}, the Iberian subset exhibits reduced deviation from the language-agnostic case due to higher proximity, while still retaining a clearly language-dependent and asymmetric structure.

\begin{table}[H]
\centering
\setlength{\tabcolsep}{6pt}
\renewcommand{\arraystretch}{1.15}\vspace{-0.2cm}
\caption{Aggregate CLTM diagnostics. Arrows indicate the direction associated with more language-agnostic behavior.}\vspace{-0.2cm}
\label{tab:baseline_cltm_diagnostics}
\resizebox{\columnwidth}{!}{%
\begin{tabular}{c c c c c}
\hline
 & $\mathrm{RFD}_1 \downarrow$ & $\mathrm{Asym}_{\mathrm{rel}} \downarrow$ & $\overline{\cos}_{\mathrm{rows}} \uparrow$ & $\text{prop}_{+} \uparrow$ \\
\hline
Iberian (5 langs) & \textbf{1.86} & \textbf{0.71} & \textbf{0.21} & \textbf{30.0\%} \\
44-languages \cite{buitrago2026cltransfer_arxiv} & 2.97 & 1.08 & 0.61 & 8.9\%\\
\hline
\end{tabular}
}
\end{table}

\subsection{Bilingual Same-Speaker Evaluation}
\label{sec:bilingual}

As observed, the CLTM exhibits clear language dependence, consistent with previous findings for the same model \cite{buitrago2026cltransfer_arxiv, buitrago2026towards}. We next examine whether part of this degradation is reduced when restricting the evaluation to the same bilingual speakers.\vspace{-0.35cm}

\begin{figure}[H]
    \centering

    \begin{subfigure}[b]{0.49\columnwidth}
        \centering
        \includegraphics[width=\linewidth]{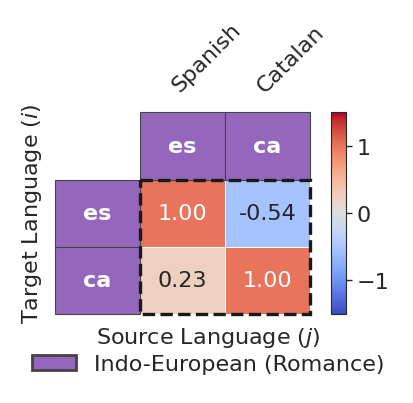}\vspace{-0.3cm}
        \caption{Standard evaluation}
    \end{subfigure}
    \hfill
    \begin{subfigure}[b]{0.49\columnwidth}
        \centering
        \includegraphics[width=\linewidth]{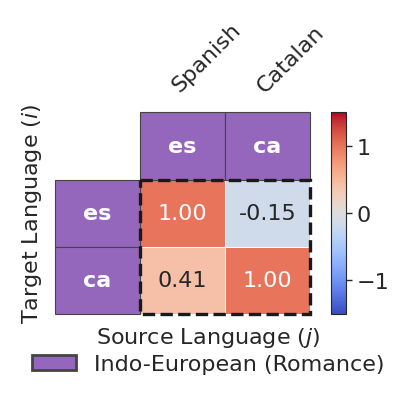}\vspace{-0.3cm}
        \caption{Same-speaker evaluation}
    \end{subfigure}\vspace{-0.3cm}
    \caption{Spanish-Catalan CLTM comparison.}
    \label{fig:es_ca}
\end{figure}\vspace{-0.7cm}

\begin{figure}[H]
    \centering
    \begin{subfigure}[b]{0.48\columnwidth}
        \centering
        \includegraphics[width=\linewidth]{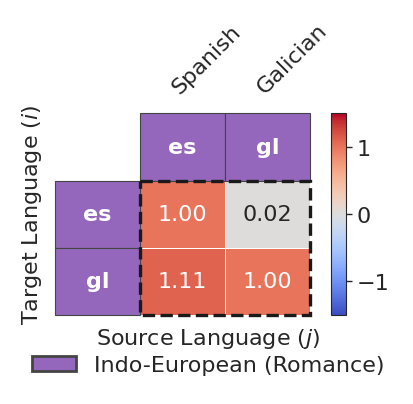}\vspace{-0.3cm}
        \caption{Standard evaluation}
    \end{subfigure}
    \hfill
    \begin{subfigure}[b]{0.48\columnwidth}
        \centering
        \includegraphics[width=\linewidth]{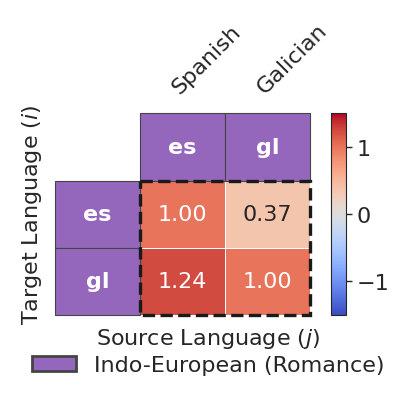}\vspace{-0.3cm}
        \caption{Same-speaker evaluation}
    \end{subfigure}\vspace{-0.3cm}
    \caption{Spanish-Galician CLTM comparison.}
    \label{fig:es_gl}
\end{figure}\vspace{-0.7cm}

\begin{figure}[H]
    \centering
    \begin{subfigure}[b]{0.48\columnwidth}
        \centering
        \includegraphics[width=\linewidth]{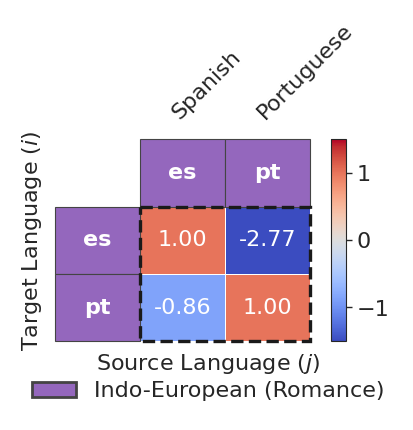}\vspace{-0.3cm}
        \caption{Standard evaluation}
    \end{subfigure}
    \hfill
    \begin{subfigure}[b]{0.48\columnwidth}
        \centering
        \includegraphics[width=\linewidth]{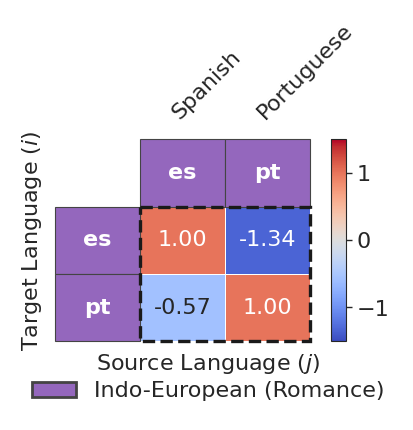}\vspace{-0.3cm}
        \caption{Same-speaker evaluation}
    \end{subfigure}\vspace{-0.3cm}
    \caption{Spanish-Portuguese CLTM comparison.}
    \label{fig:es_pt}
\end{figure}\vspace{-0.7cm}

\begin{figure}[H]
    \centering
    \begin{subfigure}[b]{0.48\columnwidth}
        \centering
        \includegraphics[width=\linewidth]{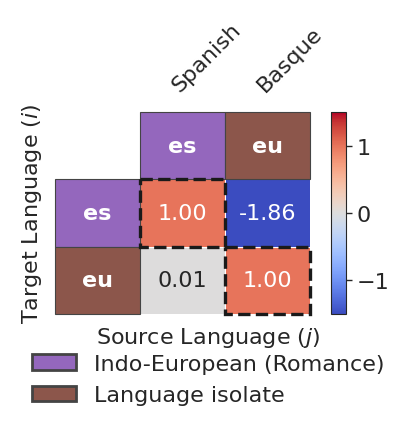}\vspace{-0.3cm}
        \caption{Standard evaluation}
    \end{subfigure}
    \hfill
    \begin{subfigure}[b]{0.48\columnwidth}
        \centering
        \includegraphics[width=\linewidth]{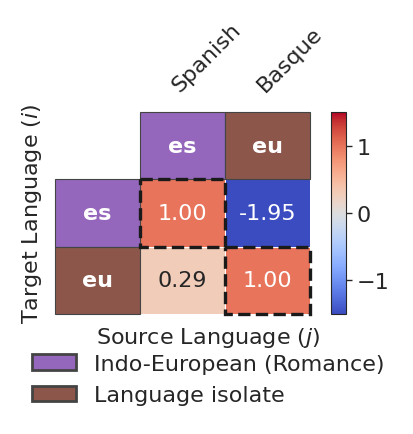}\vspace{-0.3cm}
        \caption{Same-speaker evaluation}
    \end{subfigure}\vspace{-0.3cm}
    \caption{Spanish-Basque CLTM comparison.}
    \label{fig:es_eu}
\end{figure}

A qualitative inspection of pairwise CLTMs suggests a systematic effect of speaker control. In Romance pairs, bilingual evaluation attenuates extreme off-diagonal values and yields more balanced bidirectional interactions, whereas the Basque pair retains a strong directional structure with limited changes.

To quantify these observations, we analyze the pairwise CLTMs using $\mathrm{RFD}_1$ and $\mathrm{Asym}_{\mathrm{rel}}$, as these capture, respectively, global deviation from language-agnostic transfer and directional asymmetry between languages. The remaining CLTM statistics are less informative in this reduced $2\times2$ setting.\vspace{-0.1cm}

\begin{table}[h]
\centering
\setlength{\tabcolsep}{5pt}
\renewcommand{\arraystretch}{1}
\caption{Comparison of CLTM metrics between standard and bilingual evaluation settings. Lower is better for both metrics.}\vspace{-0.2cm}
\label{tab:cltm_comparison}
\resizebox{\columnwidth}{!}{%
\begin{tabular}{cl c cc c ccc}
\toprule
&&& \multicolumn{2}{c}{$\mathrm{RFD}_1 \downarrow$}
&& \multicolumn{2}{c}{$\mathrm{Asym}_{\mathrm{rel}} \downarrow$} \\
\cmidrule(lr){4-5}
\cmidrule(lr){7-8}
&\textit{Pair} && Standard & Bilingual && Standard & Bilingual & \\
\midrule
&es-ca && 0.801 & \textbf{0.575} && 0.682 & \textbf{0.496} & \\
&es-eu && \textbf{1.411} & 1.552 && \textbf{1.154} & 1.373 & \\
&es-gl && 0.511 & \textbf{0.309} && 0.889 & \textbf{0.588} & \\
&es-pt && 2.154 & \textbf{1.469} && 0.811 & \textbf{0.545} & \\
\bottomrule
\end{tabular}%
}\vspace{-0.1cm}
\end{table}

Table~\ref{tab:cltm_comparison} shows that bilingual same-speaker evaluation has a consistent effect across language pairs. For closely related languages, such as Spanish--Catalan and Spanish--Galician, the bilingual setting consistently reduces both deviation and asymmetry, indicating that part of the previously observed cross-lingual gap is attributable to speaker variability (Fig.~\ref{fig:es_ca}, Fig.~\ref{fig:es_gl}). 

In contrast, Basque and Portuguese show a different pat-
tern. Spanish--Basque remains strongly asymmetric and even exhibits a slight increase in $\mathrm{RFD}_1$ under bilingual evaluation, and Spanish--Portuguese shows reduced but still substantial negative transfer (Fig.~\ref{fig:es_pt}), suggesting a limited effect of speaker variability and a stronger role of language interference.


\subsection{Analysis of Language-Induced Embedding Shifts}

Although restricting the evaluation to bilingual speakers attenuates language-dependent effects, it does not eliminate them. We therefore investigate whether, even for the same speaker, changing language induces systematic shifts in the embedding space.

For each bilingual speaker $s$, we compute per-language centroids by averaging all utterance embeddings in language $l$,\vspace{-0.25cm}\begin{equation}
\boldsymbol{\mu}_s^{\,l}
=
\frac{1}{N_s^{\,l}}
\sum_{i=1}^{N_s^{\,l}}
\mathbf{z}_{s,i}^{\,l},
\end{equation}
\vspace{-0.30cm}

where $\mathbf{z}_{s,i}^{\,l}$ denotes the embedding of $i$-th utterance in language $l$. The language-induced displacement is then defined as\vspace{-0.1cm}
\begin{equation}
\boldsymbol{\delta}_s^{\,i\rightarrow j}
=
\boldsymbol{\mu}_s^{\,j}
-
\boldsymbol{\mu}_s^{\,i},
\end{equation}\vspace{-0.5cm}

which represents how the speaker embedding changes when
switching from language $i$ to language $j$.

We characterize these displacements using two complementary metrics. First, the mean displacement magnitude,\vspace{-0.25cm}
\begin{equation}
M_{i\rightarrow j}
=
\frac{1}{S}
\sum_{s=1}^{S}
\left\lVert
\boldsymbol{\delta}_s^{\,i\rightarrow j}
\right\rVert,
\end{equation}\vspace{-0.3cm}

which measures the average strength of language-induced perturbation. Second, directional consistency across speakers,\vspace{-0.2cm}
\begin{equation}
C_{i \rightarrow j}
=
\frac{2}{S(S-1)}
\sum_{a<b}
\cos\left(
\boldsymbol{\delta}_a^{\,i\rightarrow j},
\boldsymbol{\delta}_b^{\,i\rightarrow j}
\right)
\end{equation}\vspace{-0.3cm}

which measures whether bilingual speakers undergo similar embedding shifts; near-zero values indicate speaker-specific effects; higher values indicate a shared language-dependent shift.\vspace{-0.15cm}

\begin{table}[H]
\centering
\setlength{\tabcolsep}{8pt}
\renewcommand{\arraystretch}{1}
\caption{Analysis of language-induced embedding shifts.}\vspace{-0.25cm}
\label{tab:geometric_analysis}
\resizebox{\columnwidth}{!}{%
\begin{tabular}{clccccc}
\toprule
 && es $\leftrightarrow$ ca & es $\leftrightarrow$ eu & es $\leftrightarrow$ gl & es $\leftrightarrow$ pt &\\
\midrule
&$M_{i\rightarrow j}$ & 0.516 & 0.538 & 0.425 & \textbf{0.785} & \\
&$C_{i\rightarrow j}$ & 0.258 & 0.248 & 0.026 & \textbf{0.421} & \\
\bottomrule
\end{tabular}%
}\vspace{-0.3cm}
\end{table}

As shown in Table~\ref{tab:geometric_analysis}, Portuguese, which previously showed the strongest cross-lingual degradation, exhibits both the largest displacement magnitude and highest directional consistency. In contrast, Galician, which previously showed minimal degradation, shows the smallest and least structured shifts. Catalan and Basque present intermediate but similar geometric patterns despite different CLTM behavior, indicating that embedding shifts alone do not directly explain cross-lingual degradation patterns.

\subsection{Linguistic Interpretation of Cross-Lingual Effects}

The CLTM patterns and embedding analysis are broadly consistent with the phonological relationships among the considered languages. 
Spanish–Galician shows the smallest embedding shift and weakest cross-lingual degradation, consistent with their high phonological proximity. Both languages share a largely overlapping consonantal inventory and a closely aligned five-vowel system \cite{a7e6813ab6bd4f28b900089546f7f583, phoible_spanish_164}, while Galician additionally allows limited mid-vowel contrasts \textipa{/E, O/} in stressed syllables \cite{regueira2010dicionario, phoible_galician_2619}.

Spanish--Catalan exhibits intermediate CLTM shifts, consistent with partial phonological overlap but increased representational mismatch relative to Spanish--Galician. Catalan expands the Spanish inventory with additional vocalic contrasts (notably \textipa{/@/} and \textipa{/y/}), and fricatives absent in Spanish (\textipa{e.g., /v, z, Z/}) \cite{a7e6813ab6bd4f28b900089546f7f583, phoible_spanish_164, phoible_catalan_1138}. These additions induce asymmetric transfer, with Spanish-trained representations generalizing more readily to Catalan than vice versa due to its richer phonological space.

Spanish–Portuguese exhibits the strongest CLTM degradation and largest embedding shifts, consistent with maximal phonological divergence in the Iberian Romance set. Portuguese expands the Spanish vowel system with schwa-like \textipa{/@/}, a set of nasal vowels (\textipa{/~i, ~e, ~o, ~u, ~@/}), unstressed vowel reduction, and additional consonantal contrasts (e.g., \textipa{/v, z, R/}), producing substantial acoustic mismatch relative to Spanish \cite{phoible_spanish_164, phoible_portuguese_163}. The CLTM asymmetry aligns with well-established intelligibility findings \cite{jensen1989}, which show that Portuguese speakers more readily comprehend Spanish than vice versa.

Spanish–Basque shows partial phonological overlap despite sharing the five-vowel system, with divergence concentrated in consonantal structure: sibilant contrasts, additional fricatives (\textipa{/z, Z, h/}), affricates (\textipa{/ts, tS/}), and stricter phonotactics \cite{phoible_spanish_164, phoible_basque_179}. This induces substantial representational strain despite vocalic alignment, with asymmetry arising as Spanish preserves the shared vowel backbone while Basque introduces consonantal contrasts poorly captured by Spanish-trained embeddings.

\section{Conclusions}

We study cross-lingual SV across five Iberian languages using a HuBERT-based model and analyze transfer patterns with the Cross-Lingual Transfer Matrix (CLTM), observing strong language-dependent behavior aligned with linguistic similarity.

A bilingual same-speaker evaluation shows that using the same speakers reduces part of the cross-lingual degradation but preserves the overall structure of language-dependent transfer.

Language switching induces systematic embedding shifts even for the same speaker, indicating that cross-lingual degradation arises from both speaker variability and intrinsic language effects, with language mismatch as the dominant factor.

\FloatBarrier
\bibliographystyle{IEEEtran}
\bibliography{Iberspeech2026_BibEntries}

\end{document}